\documentclass{amsart}
\usepackage{mathrsfs,amssymb,amsmath,amsfonts}
\usepackage{hyperref,graphicx,color}

\theoremstyle{plain}
\newtheorem{principle}{Principle}
\numberwithin{equation}{section}
\sloppypar
\begin{document}
\title[A quantum variant of the hole argument]{A quantum variant of Einstein's hole argument}
\author{I. Schmelzer}
\email{ilja.schmelzer@gmail.com}%
\urladdr{ilja-schmelzer.de}
\thanks{Berlin, Germany}
\keywords{quantum gravity, superposition}%

\begin{abstract}
We extend Einstein's hole argument into the quantum domain, and argue that quantum observables for quasiclassical superpositional states of gravitational fields require additional information to be well-defined, namely, relative positions of the gravitational fields involved in superpositional states. As a consequence, for the definition of these quantum observables we need a common background. 

The observable is a transition probability in a simple double-slit experiment with partial gravitational measurement of position. It may be easily computed in non-relativistic Schr\"{o}dinger theory with Newtonian potential. 

\end{abstract}
\maketitle

\section{Introduction}

Einstein's hole argument shows the importance of looking for observable predictions of a theory: It gives us different solutions $g_{\mu\nu}(x)$, $g'_{\mu\nu}(x)$ for the same initial values and boundary conditions. Nonetheless, the two solutions cannot be distinguished by observation. 

In this paper we argue that the situation in the quantum domain is different. We have a qualitatively new element -- the superposition of different gravitational fields. There are two different generalizations of the hole argument: One which applies the same diffeomorphism to all fields (which we name c-covariance), and one where different diffeomorphisms may be applied to different fields (q-covariance). A truly background-free theory of quantum gravity, which is based only on covariant equations, has to be q-covariant, while a c-covariant theory needs some additional structure, which connects the different solutions involved in a quasiclassical superpositional state, especially allows to define a notion of ``the same diffeomorphism'' for different solutions. Especially, the equations of GR define the involved fields only modulo diffeomorphism, and there is no well-defined notion of ``the same diffeomorphism'' for two different solutions of the Einstein equations. As a consequence, many researchers think that quantum gravity has to be a q-covariant theory (see, for example, \cite{Anandan}), or, in other words, background-free. 

Instead, we argue, that a q-covariant theory is not viable. In quantum gravity, there will be new, quantum observables, which cannot be computed in a q-covariant theory. 

At a first look, it seems a strange idea to postulate the existence of observables in a theory -- quantum gravity -- which does not yet exist. But a quite simple quantum theory of gravity already exists: non-relativistic multi-particle Schr\"{o}dinger theory with Newtonian interaction potential, or, shortly Newtonian quantum gravity (NQG). It seems reasonable to assume that NQG has to appear as the non-relativistic limit of full QG. As well, it seems reasonable to assume that observables of NQG have to be defined in full QG as observables, at least for sufficiently weak, non-relativistic gravitational fields.

Now, we can find in NQG a simple observable which depends on the relative position of different gravitational fields involved in a superpositional state, and which cannot be defined in a q-covariant theory. The new observable is a simple transition probability $\rho_{trans}$ in a variant of a double-slit experiment. It measures, if and how much of a superposition is destroyed by gravitational interaction of the state with a test particle.

This quantum observable gives a new, quantum version of the hole argument. Different from Einstein's original argument, it is based on an observable $\rho_{trans}$, which cannot be defined in a q-covariant theory. As a consequence, a q-covariant quantum theory of gravity is unable to predict the observable $\rho_{trans}$, and is, therefore, not viable. 

We consider various solutions of this problem. We favour the introduction of a common background for different gravitational fields. This solution requires an additional classical ($\hbar$-independent) physical equation, for example, the harmonic condition, which breaks general covariance and defines a common background. 

\section{The classical hole argument}

Let's remember Einstein's hole argument \cite{Einstein} and it's resolution in classical general relativity. Let $g_{\mu\nu}(x)$ a solution of the Einstein equation. Let's consider some bounded spacetime region $\Sigma$ -- the ``hole''. It is located in the future of the initial time $x^0=t_0$.
\footnote{
In Einstein's version it was located in a region without any material processes. The conclusion was, correspondingly, that the gravitational field cannot be uniquely defined by the distribution of matter. This contradicts Mach's principle. But that GR violates Mach's principle is something we have accepted. The gravitational field has it's own degrees of freedom, gravitational waves. Thus, this version of the hole argument has only historical interest.
} 
Then we consider a nontrivial smooth diffeomorphism $x' = x'(x)$, which is trivial ($x'=x$) outside $\Sigma$. As a consequence of diffeomorphism invariance, the transformed metric $g'_{\mu\nu}(x')$, after replacing $x'$ with $x$, gives a different solution $g'_{\mu\nu}(x)$ of the Einstein equations. It coinsides with $g_{\mu\nu}(x)$ outside $\Sigma$, thus, defines a different solution with the same boundary conditions and the same set of initial data at $x^0=t_0$. Thus, at a first look, it seems that in GR the gravitational field cannot be uniquely defined by whatever set of initial values and boundary conditions, for matter fields as well as the gravitational field.

The solution of this problem is that the two solutions $g_{\mu\nu}(x)$ and $g'_{\mu\nu}(x)$, while different as functions of $x$, cannot be distinguished by observation. What naively looks like an observable -- the value $g_{\mu\nu}(x)$ in a point $x$ -- is not observable. Observables are connected with events, which have to be identified by their relations to other events. For example, the event $x_1$ may be identified by the set of events at $t_0$ which intersect it's past light cone. But this same set at $t_0$ defines, on the $g'_{\mu\nu}(x)$, another event $x_1'=x'(x_1)$. 

In this sense, all real, physical observables may be, despite this argument, well-defined by the initial values and boundary conditions.

We see that for the viability of the covariant Einstein equations, it is essential that there is no observable $O$ which allows to distinguish the two solutions $g_{\mu\nu}(x)$ and $g'_{\mu\nu}(x)$ connected via the coordinate transformation $x'=x'(x)$ in the hole $\Sigma$. 

Let's note that Einstein's original assumption, that there should be no material inside the hole, can be removed. We can add an arbitrary set of material fields $\phi(x)$. Applying the hole argument, we obtain two sets of solutions $\{g_{\mu\nu}(x),\phi(x)\}$ and $\{g'_{\mu\nu}(x),\phi'(x)\}$. The argument as well as it's solution remains unchanged. 

Last not least, the argumentation can be extended without much difficulty from the purely classical gravity into semiclassical gravity, that means, quantum theories on a fixed background. It is known that this requires, because of problems related with particle creation, the consideration of quantum field theory. Despite this, for the purpose of this paper, a single particle approximation is sufficient. As well, for the purpose of this paper purely spatial diffeomorphisms, which do not change the time variable, are completely sufficient. Therefore we do not have to consider problems in semiclassical gravity related with the choice of the time variable. Thus, we fix some foliation $x = ({\bf x},t)$ and restrict the argumentation to diffeomorphisms which preserve this foliation. That means, we describe a quantum state on a curved background $g_{\mu\nu}({\bf x},t)$ in a single particle approximation with some wave function $\psi({\bf x},t)$. The semiclassical version of the hole argument is, then, described by two sets $\{g_{\mu\nu}({\bf x},t),\psi({\bf x},t)\}$ and $\{g'_{\mu\nu}({\bf x},t),\psi'({\bf x},t)\}$ of solutions. Again, the argument can be resolved, as in the purely classical domain, by recognizing that all physical observables coinside.

\section{Quasiclassical superposition of gravitational fields}

Let's consider now a next step in the direction of quantum gravity -- a simple  superposition of two semiclassical states. That means, to describe this state, a single classical gravitational field (as in semiclassical theory) is no longer sufficient. We have to consider two semiclassical field configurations $\{g^1_{\mu\nu}({\bf x},t),\psi_1({\bf x},t)\}$ and $\{g^2_{\mu\nu}({\bf x},t),\psi_2({\bf x},t)\}$. We may write them in the Dirac notation as $|g_1,\psi_1\rangle$ and $|g_2,\psi_2\rangle$. Now we can denote our superpositional state as
\begin{equation}
	|\Psi\rangle = |g_1,\psi_1\rangle + |g_2,\psi_2\rangle.
	\label{eq:super}
\end{equation}
The classical principle of covariance gives:

\begin{principle}[simple covariance] The states $|g,\psi\rangle$ and $|g',\psi'\rangle$ cannot be distinguished by observation.
\end{principle}

But now we have already two different classical gravitational fields in a single superpositional state. How to extend the principle of covariance to such quantum superpositions? There are two possibilities:

\begin{principle}[c-covariance] The states $|g_1,\psi_1\rangle + |g_2,\psi_2\rangle$ and $|g_1',\psi_1'\rangle + |g_2',\psi_2'\rangle$ cannot be distinguished by observation.
\end{principle}

Here, we apply the same transformation $x'(x)$ to above semiclassical fields. But there is also another possibility. Let's consider two different transformations $x'(x)$ and $x''(x)$ for our two semiclassical configurations. We obtain another, more restrictive covariance principle:

\begin{principle}[q-covariance] The states $|g_1,\psi_1\rangle + |g_2,\psi_2\rangle$ and $|g_1',\psi_1'\rangle + |g_2'',\psi_2''\rangle$ cannot be distinguished by observation.
\end{principle}

Obviously, q-covariance is stronger than c-covariance: We obtain c-covariance from q-covariance for the particular case $x'(x) = x''(x)$. 

Which of the two generalizations is the appropriate one? The answer will be given, of course, only by quantum gravity. But already now it is possible to classify quantum theories of gravity based on these principles: In a really background-independent theory it is impossible to define the notion ``the same diffeomorphism'' for two different solutions. This can be done only if we have two solutions given in the same set of coordinates $x$. But the GR equations do not define this particular system of coordinates. Thus, even the notion of c-covariance is not defined in GR. Thus, the only appropriate choice for a GR-based theory of quantum gravity seems to be q-covariance. In other words, q-covariance can be considered as another word for background independence of a theory of gravity. 

Instead, if we have a theory of gravity with a fixed background, we can define the position relative to the background. This allows to define a notion of ``the same diffeomorphism'' for different gravitational fields. The background itself is not a q-covariant notion. Thus, for a theory with background, the notion of q-covariance is not natural at all. On the other hand, it is well-known, that covariant formulations are possible even for non-covariant classical theories like SR or Newtonian theory. The notion of c-covariance seems to be an appropriate generalization of such an artificial, trivial notion of covariance into the quantum domain. Thus, it seems, we can identify c-covariance of a theory of gravity with background-dependence. 

Last not least, let's note that the background does not need to have a predefined geometry. Especially an affine structure would be completely sufficient. For example, introducing the harmonic gauge is sufficient to fix a common background, so that the hole argument no longer works, but it does not define a geometry on the background. 

\section{Gravitational partial position measurement}

Let's consider now a simple thought experiment in the domain of quasiclassical quantum gravity. We have one particle, called ``source particle'', in a superpositional state of type $\delta(x-x_l) + \delta(x-x_r)$. This state interacts gravitationally with a second particle, the ``test particle''. Then, the test particle will be simply ignored, and we consider the resulting state of the source particle. There are two limiting cases: If the gravitational interaction is sufficiently strong, the final position of the test particle is a measurement of the position of the source particle. As a consequence, the interference is destroyed and the effective state of the source particle has to be described by a density matrix. If the interaction is too weak, the final position of the test particle does not depend on the position of the source particle, and for the state of the source particle nothing has changed. In intermediate situations we will find a partial destruction of the superposition.

The different outcomes of the interaction can be distinguished by subsequent observation of the source particle alone. For example, if the source particle is part of a classical double slit experiment, we can observe the presence or absence of an interference pattern. 

Let's see now how to compute the remaining ``degree of interference''. Let's denote the initial state of the source particle as $|\phi_+\rangle$, where

\begin{equation}
	|\phi_\pm\rangle = \frac{1}{\sqrt{2}}(|\phi_l\rangle \pm |\phi_r\rangle),
	\label{eq:Psi}
\end{equation}

and $|\phi_{l/r}\rangle\approx |\delta(x-x_{l/r})\rangle$ denotes states where the source particle is located near the position $x_{l/r}$ (say, the left or right slit of a double slit experiment). The test particle is prepared initially in the state $|\psi_0\rangle$. Thus, the two-particle system is prepared in the state 

\begin{equation}
	|\Psi_{in}\rangle = |\phi_+\rangle\otimes|\psi_0\rangle
	\label{eq:Psi_in}
\end{equation}

The gravitational interaction is diagonal in the particle positions. We assume that the mass of the source particle $M$ is much greater than the mass $m$ of the test particle. Thus, in some approximation the interaction leaves the position of the source particle unchanged, and changes only the state of the test particle. This gives, for the position $x_{l/r}$ of the source particle, the wave functions $\psi_{l/r}(x)$. At the end of the interaction, we have obtained the state

\begin{equation}
	|\Psi_{out}\rangle = \frac{1}{\sqrt{2}}(|\phi_l\rangle\otimes|\psi_l\rangle + |\phi_r\rangle\otimes|\psi_r\rangle),
	\label{eq:Psi_out_lr}
\end{equation}

Now, in the final measurement we measure the eigenstates $|\phi_\pm\rangle$ of the source particle. The test particle is ignored. (That means, we assume its position is measured, but ignore it, computing only the trace.) The ``degree of loss of interference'' is defined by the probability $\rho_{trans}$ of the source particle being observed in the state $\phi_-$, and equals

\begin{equation}
	\rho_{trans} =  \frac{1}{2}(1 - \Re \langle \psi_l|\psi_r\rangle).
	\label{eq:rho}
\end{equation}

Especially, if there is no interaction, with $|\psi_l\rangle=|\psi_r\rangle$, we have $\rho_{trans}=0$. In the case of complete measurement, which corresponds to no interference pattern, $\langle\psi_l|\psi_r\rangle=0$, thus, $\rho_{trans}=\frac{1}{2}$. We conclude that (at least) the real part of the scalar product $\langle \psi_l|\psi_r\rangle$ is observable.

\subsection{Description in terms of Newtonian quantum gravity} 

Let's see now how to compute the scalar product $\langle\psi_l|\psi_r\rangle$ in Newtonian quantum gravity. We use the approximation $\phi_{l/r}(x) = \delta(x-x_{l/r})$. In this approximation, the two-particle problem reduces to two one-particle problems with a classical source of the gravitational field in $x_{l/r}$.  Thus, we have to solve only the two one-particle Schr\"{o}dinger equations

\begin{equation}
	i\partial_t \psi_{l/r}(x,t) = (-\frac{1}{2m} \Delta - \frac{mM}{|x-x_{l/r}|})\psi_{l/r}(x,t)
	\label{eq:Schroedinger}
\end{equation}

for the initial value $\psi_{l/r}(x,t_0) = \psi_0(x)$. Then we can compute the transition probability $\rho_{trans}$ by the well-defined expression 

\begin{equation}
	\rho_{trans} = \frac{1}{2}(1 - \Re \int \overline{\psi}_l(x)\psi_r(x) d^3x)
	\label{eq:transNQG}
\end{equation} 

\subsection{Description in terms of semiclassical general relativity}

Now, we would like to compute the first GR corrections for our well-defined NQG observable $\rho_{trans}$. 

There appear, of course, the usual problems of one-particle theory in the relativistic domain and all the other problems in the domain of semiclassical quantum theory. But these problems are solvable using the standard techniques of semiclassical QFT. Let's therefore assume that, in an appropriate approximation (which ignores particle creation and so on), using for simplicity a fixed foliation, we can find, for a given gravitational field $g_{\mu\nu}({\bf x},t)$, a corresponding one-particle wave function $\psi({\bf x},t)$. 

Then, we use the Schwarzschild solutions $g^{l/r}_{\mu\nu}({\bf x},t)$ for the source particle located in $x_{l/r}$ and obtain two pairs of solutions $\{g^l_{\mu\nu}({\bf x},t),\psi_l({\bf x},t)\}$ and $\{g^r_{\mu\nu}({\bf x},t),\psi_r({\bf x},t)\}$. As far, everything seems nice. It remains to compute the scalar product. The straightforward formula would be
\begin{equation}
	\rho_{trans} = \frac{1}{2}(1 - \Re \int \overline{\psi}_l(x)\psi_r(x)d\mu),
	\label{eq:transGR}
\end{equation} 
where we leave the question of the definition of the measure $d\mu$ open. But now let's consider the covariance properties of our expression for $\rho_{trans}$. If we have correctly managed the transformation rules for $\psi$ and $d\mu$ for changes of coordinates, we obtain, without problems, weak covariance:

\begin{equation}
	\int \overline{\psi}_l(x)\psi_r(x) d\mu = \int \overline{\psi}'_l(x)\psi'_r(x) d\mu'
	\label{eq:weakPsi}
\end{equation}

But strong covariance fails completely. For example, assume $\psi_l(x)=\psi_r(x)$, thus, $\rho_{trans} = 0$, with an ideal, unchanged interference picture. Assume that $\psi_l(x)$ has finite support $U$. Now consider the deformations $x'=x'(x), x''=x''(x)$ so that $U'\cap U''=\emptyset$. As a consequence, $\int \overline{\psi}'_l(x)\psi''_r(x)d\mu = 0$ (whatever the measure $d\mu$), and we obtain
\begin{equation}
	\tilde{\rho}_{trans} = \frac{1}{2}(1 - \Re \int \overline{\psi}_l(x')\psi_r(x'')d\mu) = \frac{1}{2} \neq \rho_{trans},
	\label{eq:transFalse}
\end{equation} 
thus, no interference picture at all. We conclude that our NQG observable $\rho_{trans}$ is not strong covariant. 

Let's remember now that the equations of general relativity define the metric $g_{\mu\nu}$ only modulo arbitrary coordinate transformation. So, we simply don't know, which of the two choices, $\overline{\psi}_l(x)\psi_r(x)d\mu$, or $\overline{\psi}_l(x')\psi_r(x'')d\mu$, we have to use to define $\rho_{trans}$.  Thus, the NQG observable $\rho_{trans}$ cannot be computed using only the equations of GR, together with semiclaccial quantum theory on a fixed GR background.

In geometric language, the two different solutions $\{g^l_{\mu\nu}({\bf x},t),\psi_l({\bf x},t)\}$ and $\{g^r_{\mu\nu}({\bf x},t),\psi_r({\bf x},t)\}$ of semiclassical theory live on different manifolds. The scalar product of functions defined on different manifolds is undefined. To define it, we need an additional object -- a map between the two manifolds. 

\section{Consequences}

As the consequence of this consideration, we conclude, that a strong covariant theory of gravity (in other words, a background-independent quantum theory of gravity) cannot have a correct NQG limit, and, therefore, is not viable.  

Let's consider some possible objections. 

First, maybe in full quantum gravity $\rho_{trans}$ is not an observable, but becomes observable only in the limit? But, whatever the observables of full quantum gravity, we should be able to derive the observables we can observe in the domain of application of NQG. Note that, at least in some sense, we already today can and do observe $\rho_{trans}$. Indeed, we observe interference patterns. Now, the particles, which show the interference patterns, always interact gravitationally with all other particles. Of course, this interaction is far too small to lead to different functions $\psi_{l/r}(x)$. That's why the NQG prediction gives $\rho_{trans}=0$. This prediction is in full agreement with the fact that we can observe interference patterns. But, even if the prediction $\rho_{trans}=0$ is quite trivial, the fact that there is agreement with observation even today is a quite nontrivial one, especially given that $\rho_{trans}$ is completely undefined in a fully covariant theory. Note also that the difference between the two results $\rho_{trans}$ and $\tilde{\rho}_{trans}$ is very big, as big as possible. In this sense, $\rho_{trans}$ is not only observable, but easily observable. 

Note also that the problem with the computation of $\rho_{trans}$ is not a problem for strong, relativistic gravitational fields.  The problem appears whenever we consider a superposition of gravitational fields. Even for gravitational fields which can be, as accurate as you like, approximated by a Newtonian potential. And, at least in principle, even if the gravitational field is as weak as you like. It is, even in this limit, possible to chose $x'=x'(x), x''=x''(x)$ so that $\int \overline{\psi}'_l(x)\psi''_r(x)d\mu$ becomes zero, in fatal and obvious disagreement with the observation of interference patterns.

But, then, maybe Newtonian quantum gravity is not a correct limit of full quantum gravity? But even this does not help. Last not least, the conceptual problem remains the same even if we consider the non-gravity limit of NQG, which is simply non-relativistic quantum mechanics. Even in this limit we obtain $\rho_{trans}=0$ as the value for transitions caused by gravity, and even in this limit we can compare with observation and obtain full agreement. Moreover, some experiments (with energy levels of neutrons in the gravitational field of the Earth), which require Newtonian quantum gravity, have been already done, and they have shown agreement with the (quite simple) predictions \cite{neutrons}. 

\subsection{The alternative: A preferred background}

The simple way to fix this problem is to fix a preferred background. This can be done using a gauge condition which fixes a preferred system of coordinates. The rule for the computation of $\rho_{trans}$ is, then, the following: We have to transform the solutions $\{g^{l/r}_{\mu\nu},\psi_{l/r}\}$ into the preferred coordinates, so that $\psi_{l}({\bf x},t)=\psi_{r}({\bf x},t)$ before the interaction $t<t_0$. Then we can use the wave functions $\psi_{l/r}$, in the preferred coordinates, to compute the scalar product $\langle\psi_l|\psi_r\rangle$.

This prescription obviously depends on the choice of the gauge condition. Different choices of the gauge condition lead to different predictions for $\rho_{trans}$. It seems, at least in principle, possible to construct a sufficiently artificial coordinate condition such that this prescription leads to $\rho_{trans}\neq 0$ even for the usual interference patterns we observe today. (Of course, such a condition has to be very strange, especially it has to define for two metrics which are very close to each other very different preferred coordinates. But in principle this would be possible.) Thus, the choice of the gauge condition, in combination with the prescription for the computation of $\rho_{trans}$ above, is a physical choice. The gauge condition is, in this sense, a physical equation, no longer an arbitrary human choice. Therefore, it seems more appropriate to name it an equation. Once it defines preferred coordinates, and ``coordinates'' different solutions, we suggest to name it ``coordinate equation''.

Now, once it is clear that we need a new physical equation, which fixes a preferred system of coordinates, we have to postulate this equation. The harmonic coordinate equation 
\begin{equation}
	\partial_\mu (g^{\mu\nu}(x)\sqrt{-g}) = 0
	\label{eq:harmonic}
\end{equation}
seems to be a favourable choice for an additional physical equation. In harmonic coordinates, the hole problem disappears. Indeed, harmonic coordinates are quite appropriate for the initial value problem, so that local uniqueness can be proven \cite{Choquet-Bruhat} \cite{Choquet-Bruhat1}. 

The harmonic condition may be introduced as an additional physical equation into GR. But in this case, we no longer have a Lagrange formalism for all equations of the theory. If we want to obtain the harmonic equation as an Euler-Lagrange equation, we have to add some gauge-breaking term, which also modifies the Einstein equations themself. There are two alternative metric theories of gravity following this way: First, there is the ``relativistic theory of gravity'' (RTG) \cite{Logunov, Logunov1}, a theory of gravity with massive graviton, which gives the Einstein equation in the limit of zero graviton mass $m_g\to 0$. Then there is the ``general Lorentz ether theory'' (GLET) \cite{Protvino, glet, clm}, with two additional parameters $\Xi, \Upsilon$, which also gives the Einstein equation for $\Xi, \Upsilon \to 0$. The main differences between these two alternatives are different metaphysics. RTG is a bimetric theory on a Minkowski background. Instead, GLET is an ether theory with classical Newtonian spacetime as the background.

A fixed Minkowski background, which allows to compare different gravitational fields, and, therefore, to compute $\rho_{trans}$, exists also in string theory.  

\subsection{Is a preferred background necessary?}

While a preferred background solves the problem with the computation of $\rho_{trans}$, it seems natural to ask if such a background is the only way to solve the problem of defining $\rho_{trans}$. 

Let's note that our considerations work only for gravitational fields which may be superposed. There may be superrules which forbid some superpositions, so that all gravitational fields split into classes of fields which may be superposed with each other. We do not consider this possibility. Thus, we assume that all gravitational fields may be superposed with each other. 

In this case, let's fix a single gravitational field, say, the vacuum metric $\eta_{\mu\nu}$, as a reference field. Let's, then, take another, arbitrary, field $g_{\mu\nu}(y)$. Then, consider pairs of solutions $\{\eta_{\mu\nu}(x),\varphi(x)\}$ and $\{g_{\mu\nu}(y),\psi(y)\}$. Assume that the transition probability 
\begin{equation}
	\rho_{trans} = \frac{1}{2}(1 - \Re \langle\varphi|\psi\rangle).
	\label{eq:trans}
\end{equation} 
is well-defined. Then we can, obviously, use  $\varphi(x)=\delta(x-x_0)$ defined on the on $\eta_{\mu\nu}$ background as a test function and obtain for real-valued functions $\psi(y)$ on the curved background, a corresponding real-valued function $\tilde{\psi}(x_0)$
\begin{equation}
	\tilde{\psi}(x_0) = \Re \langle\delta(x-x_0)|\psi\rangle = 1 - 2 \rho_{trans}.
	\label{eq:test}
\end{equation}
In other words, we can compute a corresponding representation of the wave function $\tilde{\psi}(x)$ on the flat background. In the simplest case, this representation is simply defined by some map $y(x)$ of the underlying spaces, so that $\tilde{\psi}(x) = \psi(y(x))$, or
\begin{equation}
	\langle\varphi|\psi\rangle = \int \overline{\varphi}(x)\psi(y(x))d^3x.
	\label{eq:scalarProduct}
\end{equation} 
In principle, there may be more general situations: The measurement of the background position on $\eta_{\mu\nu}$ may be incompatible with the position measurement on the curved background. But even in this case, the measurement of the background position on a common reference background $\eta_{\mu\nu}$ is well-defined, with the corresponding wave function $\tilde{\psi}(x)$.

\section{One possibility to save the background-independent approach}

Let's consider shortly one idea, which, possibly, could be tried out to save the background-independent approach.  Instead of considering only the interaction between the source particle and the test particle, let's consider the whole  double slit experiment, from the beginning to the end, in it's most classical form, that means, with a common starting point for the source particle, and a common final point of the source particle on the screen. In this case, as in the initial state, as in the final state, we have no longer superpositional states of gravitational fields. Thus, the superpositional hole argument cannot be used, once no superposition is present. 

Assume all is fine with this argument. What follows? Even in our simple double slit experiment, the splitting can be done as far in the past as we like. And the final measurement can be done as far in future as we like. If we have to trace back all superpositional states until we can reduce them in such a way to a common, non-superpositional origin, as well as to a superposition-free final state, we will end with some sort of S-matrix-like theory, which does not allow us to compute anything for finite distances. 

Now, S-matrix theory is, without doubt, of high value for pragmatical computations for scattering experiments in particle accelerators. But we should clearly reject the idea that a fundamental theory of everything may be such an S-matrix theory. Even for scattering experiments in particle accelerators the S-matrix gives only a fixed level of accuracy of the predictions. For more accurate predictions, it becomes necessary to compute corrections related with the finite size of our measurment devices.  A theory which, even in principle, is unable to do this, is not viable as a theory of the universe. Moreover, an S-matrix approach is obviously completely useless for cosmological consideration. What we observe, we observe from a very tiny part of the universe, our Solar system. Thus, a theory of everything should be able to compute predictions for finite distances, in space as well as in time. A theory, which cannot do it in principle, is simply not acceptable as a theory of the universe. The aim of the search for quantum gravity and theories of everything is not the computation of S-matrices: These can be computed with high enough accuracy without them. The SM is sufficient for this purpose. The aim of such theories is to obtain a common conceptual scheme for everything in the universe. And a theory which gives only S-matrices, obviously, fails to give this scheme for our local, human observations.

Thus, we need a theory which allows to compute predictions in finite space and time. And a quantum theory of gravity which describes what happends here and now cannot live without superpositions of gravitational fields. The idea to trace back such superpositions until we find a superposition-free state in the past should, therefore, be rejected. 

Against attempts to deny our new observable the right to exist in full quantum gravity we can, last not least, propose a general argument: They do not take the lecture teached by quantum theory seriously. Our new observable $\rho_{trans}$ is a new, purely quantum, observable. In classical theory, there is nothing comparable: There are no superpositions which may be destroyed by observation. Such qualitatively new observables also define some deep insight into nature. And we should not ignore this lecture, should not close our eyes, but look at these new, quantum observables, try to find out if they allow us to see more. 

Note also that nothing of the beauty of the classical symmetry is lost, if we find that the symmetry in the quantum case is different, if we find, that quantum observables give us, sometimes, more and different information. The qualitatively new, quantum observable we have found in this paper allows us to distinguish states which would have been indistinguishable using only the old, classical observables. They allow us to see the common background -- an object which is hidden from observation as long as we can use only classical observables.

\section{Discussion}

Our argument suggests that q-covariant, background-independent approaches to quantum gravity are not viable. But, of course, the history of various impossibility theorem (say, von Neumann's impossibility theorem for hidden variable theories, which has been shown to be nonsensical by pilot wave theories) as well as the history of Einstein's original hole argument suggests, that we should not take impossibility arguments of the sort given here too serious. Therefore, it should not be expected that quantum gravity researchers give up their hope for a background-free quantum theory of gravity.

Moreover, I suspect that they do not find the argument very impressive. Last not least, the definition of the observables is known to be a complicate issue in quantum gravity. For example, Smolin \cite{Smolin} notes that ``\ldots one cannot define the physical observables of the theory without solving the dynamics'', Thiemann \cite{Thiemann} writes ``\ldots one must find a complete set of Dirac observables (operators that leave the space of solutions invariant) which is an impossible task to achieve even in classical general relativity.'' 

One reason for the attractiveness of the background-independent approach is, obviously, it's philosophical background. It goes back to the position of Leibniz, who has proposed arguments for a relational view, against an absolute notion of space and time proposed by Newton. A nice introduction, from point of view of the modern background-independent approach, can be found in \cite{Smolin}:

\begin{quote}
``Leibniz's argument for relationalism was based on two principles, which have been the focus of many books and papers by philosophers to the present day. The \emph{principle of sufficient reason} states that it must be possible to give a rational justification for every choice made in the description of nature. \ldots A theory that begins with the choice of a background geometry, among many equally consistent choices, violates this principle.\ldots

One way to formulate the argument against background spacetime is through a second principle of Leibniz, \emph{the identity of the indiscernible}. This states that any two entities which share the same properties are to be identified. Leibniz argues that were this not the case, the first principle would be violated, as there would be a distinction between two entities in nature without a rational basis. If there is no experiment that could tell the difference between the state in which the universe is here, and the state in which it is translated 10 feet to the left, they cannot be distinguished. The principle says that they must then be identified. In modern terms, this is something like saying that a cosmological theory should not have global symmetries, for they generate motions and charges that could only be measured by an observer at infinity, who is hence not part of the universe.''
\end{quote}

Now, we believe that relationalism is inherently wrong, because it is based on a wrong, positivistic understanding of scientific knowledge, and should be rejected. Following Popper \cite{PopperCR, PopperLSD}, we believe into the priority of theory: Theories are free guesses about Nature, not bound by principles following from observations. Based on the theory we derive what is observable as well as predictions about these observables. Only after this, observations become important, as a method to falsify some of the theories. Instead, relationalism has to be rejected on the same grounds as pre-Popperian logical positivism, as a variant of the priority of observation.

But this paper we would like to finish with a completely different argument. Namely, the argumentation against a background based on relationalism completely fails, if we take into account our new, quantum observable. Indeed, this new observable gives, as we have shown, sufficient reason for the introduction of a common background: The background solves the problem of computing a prediction for the new observable, and there is no obvious alternative way to solve this problem. It is also no longer possible to apply the principle of ``identity of the indiscernible'' against the background. Indeed, superpositional states with different values for $\rho_{trans}$ are no longer indiscernible. 

And, last not least, arguments in favour of relationalism in general fail to support the background-independent approach: The background, as constructed here, defines itself a \emph{relation} between physical objects -- a relation between two gravitational fields, which are part of one superpositional state.

\end{document}